\documentclass[12pt,aps]{revtex4}

\usepackage{graphicx}
\usepackage{bbm}
\usepackage{amssymb}
\usepackage{amsmath}

\newcommand{\nc}{\newcommand}
\newsavebox{\LSIM}
\sbox{\LSIM}{\raisebox{-1ex}{$\ \stackrel{\textstyle<}{\sim}\ $}}

\newsavebox{\GSIM}
\sbox{\GSIM}{\raisebox{-1ex}{$\ \stackrel{\textstyle>}{\sim}\ $}}

\nc{\be}{\begin{equation}}
\nc{\ee}{\end{equation}}
\nc{\bea}{\begin{eqnarray}}
\nc{\eea}{\end{eqnarray}}
\nc{\nn}{\nonumber}
     
\nc{\acom}[2]{ \left\{ #1,#2 \right\} }
\nc{\com}[2]{ \left[ #1,#2 \right] }

\nc{\lp}{\left(}
\nc{\rp}{\right)}

\nc{\rme}{{\textrm{e}}}       

\nc{\eq}{{Eq.}}
\nc{\eqs}{{Eqs.}}

\begin{document}


\title{Production of Gravitational Waves in the nMSSM}

\author{Stephan J. Huber$^{(1)}$ and Thomas Konstandin$^{(2)}$}

\email[]{s.huber@sussex.ac.uk}
\email[]{konstand@kth.se}

\affiliation{$(1)\,$ Department of Physics and Astronomy, University of Sussex,
        Brighton, BN1 9QJ, UK}

\affiliation{$(2)\,$Department of Theoretical Physics, Royal Institute of Technology (KTH), 
        AlbaNova University Center, Roslagstullsbacken 21, 106 91
        Stockholm, Sweden}

\date{\today}

\begin{abstract}
During a strongly first-order phase transition gravitational waves are
produced by bubble collisions and turbulent plasma motion. We analyze
the relevant characteristics of the electroweak phase transition in
the nMSSM to determine the generated gravitational wave
signal. Additionally, we comment on correlations between the
production of gravitational waves and baryogenesis. We conclude that
the gravitational wave relic density in this model is generically too
small to be detected in the near future by the LISA experiment.  We
also consider the case of a "Standard Model" with dimension-six Higgs
potential, which leads to a slightly stronger signal of gravitational
waves.
\end{abstract}

\maketitle

%
%

\section{Introduction}

Presently, several experiments are under consideration that could
detect for the first time a stochastic background of gravitational
waves (GWs). One of the main motivations to pursue these experiments
is that the discovery of a relic gravitational background would be a
smoking gun signal from inflation~\cite{Turner:1996ck} and hence
might allow to test the paradigm of an era of exponential expansion in
the early Universe.

Another source of stochastic GWs are strongly first-order phase
transitions. Space-based experiments, such as
LISA~\cite{Danzmann:2003tv} and BBO~\cite{Corbin:2005ny} will have
remarkably good sensitivity at frequencies that coincide with the
redshifted spectrum of GWs produced during an electroweak phase
transition at temperatures $T\sim 100$ GeV. This opens the possibility
to infer information about the electroweak phase transition from GW
observations. A strongly first-order phase transition not only would
produce GWs, but could also generate the observed baryon asymmetry of
the Universe (BAU) through electroweak baryogenesis
\cite{Kuzmin:1985mm, Anderson:1991zb}. This way the observation of a stochastic
background of GWs might teach us about electroweak baryogenesis in
extensions of the Standard Model (SM).

During a first-order phase transition there are two distinct
mechanisms that produce GWs: The colliding phase boundaries
\cite{Kosowsky:1991ua,Kosowsky:1992rz,Kosowsky:1992vn,Kamionkowski:1993fg},
and the turbulent motion of the plasma
\cite{Kosowsky:2000rq,Kosowsky:2001xp,Dolgov:2002ra,Caprini:2006jb,Gogoberidze:2007an}. 
Concerning the latter, there is an ongoing discussion in the
literature on how to correctly model the temporal correlation of the
turbulent plasma.  In particular, the different approaches predict
different peak frequencies of the resulting GW spectrum.

Studies of GWs from the electroweak phase transition have been
performed in Refs.~\cite{Nicolis:2003tg} and~\cite{Grojean:2006bp}
without relying on particular particle physics models. The results are
presented as functions of the two main parameters of the problem,
which are the typical size of the colliding bubbles and the available
energy.  It was concluded that a sufficiently strong phase transition
could lead to an observable GW signal at LISA. Of course, in a given
model the bubble size and the available energy are linked, and it is
interesting so see what happens to the GW signal. The cases of the
MSSM and the NMSSM (Next-to-minimal supersymmetric SM) were studied in
Ref.~\cite{Apreda:2001us}.  While it was found that in the MSSM the
produced amount GWs is orders of magnitude below the LISA sensitivity,
the situation in the NMSSM seems much more promising.  The cubic terms
present in the NMSSM tree-level Higgs potential can lead to a much
stronger phase transition \cite{NMSSM}.

However, the authors of Ref.~\cite{Apreda:2001us} used a crude method
to determine the bubble configurations, which overestimates the
strength of the phase transition and the GW signal. The aim of this
work is to improve on this point to arrive at a more realistic
estimate of the signal strength. To achieve this we apply our recently
presented numerical method to compute the bubble shapes
\cite{Konstandin:2006nd}.

Like in Ref.~\cite{Apreda:2001us} we consider a singlet extension of
the MSSM, in our case the nearly Minimal Supersymmetric Standard Model
(nMSSM) \cite{nMSSM}. In contrast to the models studied in
Refs.~\cite{Apreda:2001us,NMSSM}, it solves the $\mu$-problem of the
MSSM, without creating a domain wall problem or destabilizing the
electroweak hierarchy through divergences of the singlet tadpole. With
this motivation, the cosmology of the nMSSM has been been studied
intensively in the literature. It was shown in
Ref.~\cite{Menon:2004wv} that the model can have a strong enough phase
transition for baryogenesis, while the lightest neutralino can at the
same time provide the dark matter of the Universe. The baryon
production was studied with a positive result recently in
Ref.~\cite{Huber:2006wf}.  Finally, Ref.~\cite{Balazs:2007pf}
investigated to what extent these cosmological issues can be studied
at colliders, in particular at the LHC.

In this paper, we compute the GW signal in the nMSSM using numerically
determined bubble configurations. In addition, correlations between
the amount of the produced GWs and the generated BAU are
discussed. Our results for GW production are less optimistic than the
estimates of Ref.~\cite{Apreda:2001us}, making it doubtful for LISA to
detect a GW signal, even for an optimal choice of parameters. BBO, on
the other hand, could detect GWs if the nMSSM phase transition is
extremely strong.  This study is a continuation of work on the baryon
asymmetry in the nMSSM that has been published recently in
Ref.~\cite{Huber:2006wf}. Since the formalism to determine the BAU,
the analysis of the phase transition and phenomenological aspects of
the nMSSM are discussed in detail in Refs.~\cite{Huber:2006wf,
Menon:2004wv}, we will be rather brief on these issues. We also study
GW production in the SM augmented by dimension-six operators.  Such a
model can be viewed as the effective low energy description of some
strongly coupled dynamics at the TeV scale. As this model has only one
Higgs field and few parameters, the analysis is much simplified and
more transparent compared to the nMSSM case.  It has been shown that
an $H^6$-term in the Higgs potential easily leads to a strong phase
transition~\cite{Z93,GSW04,BFHS04}. Additional dimension-six operators
can also provide new sources of CP violation that can account for the
observed baryon asymmetry~\cite{BFHS04,FH06} without generating large
electric dipole moments~\cite{HPR06}.

The paper is organized as follows. The next section contains the
formalism to determine the relevant properties of the phase
transition. In Sec.~\ref{sec_phi6} the $H^6$ model is discussed, while
in Sec.~\ref{sec_model} the nMSSM and its temperature dependent
effective potential is summarized. In Sec.~\ref{sec_num} the numerical
results for the nMSSM are presented before we conclude in
Sec.~\ref{sec_con}. Appendix~\ref{sec_PT} contains a brief review of
first-order phase transitions in cosmology.

\section{Gravitational Waves\label{sec_GW}}

The discussion of GW production in this section closely follows the
analysis in Ref.~\cite{Nicolis:2003tg}. Details about the GW
production during a first-order phase transition can be found in this
work and the references therein.

One source of GWs during the phase transition results from colliding
Higgs bubbles at the end of the phase transition.  Depending on the
strength of the phase transition, the bubble wall profile propagates
faster or slower than the speed of sound, which is $1/\sqrt{3}$ in a
relativistic thermal bath. In the former case GWs are produced by {\it
detonation}, while in the latter case {\it deflagration} is the
dominating process. Since GW production is strongly suppressed in a
deflagration, we focus in the following on the detonation mode as
discussed in
Refs.~\cite{Steinhardt:1981ct,Kosowsky:1991ua,Kosowsky:1992rz,Kosowsky:1992vn,Kamionkowski:1993fg}.

To determine the magnitude of the produced GWs, two parameters of the
phase transition are essential. The first one is related to the latent
heat,
\be
\label{def_lat_heat}
\epsilon_* = -\Delta V + T_* \left. \frac{\partial V}{\partial T} \right|_{T_*},
\ee
which provides the energy available to be transferred to GWs. The
amount of GWs depends on the ratio of the latent heat to the energy
density of the radiation in the plasma,
\be
\label{alpha_def}
\alpha = \frac{30 \epsilon_*}{\pi^2 g_* T_*^4}.
\ee
This parameter measures the strength of the phase transition, as
relevant for the GW production.  $T_*$ denotes the temperature of the
phase transition, and $g_*$ is the effective number of degrees of
freedom (in the SM: $g_*=107.75$). There are several prescriptions to
define the start and end of the phase transition, and the time of
maximal GW production. This temperature is further specified in the
Appendix, see \eq~(\ref{def_Tf}).

The second parameter that enters the production of GWs is the typical
radius of the colliding bubbles, $\left< R \right>$.  It sets the
length scale of the problem. It can roughly be expressed as the
product of the wall velocity, $v_b$, and the duration of the phase
transition, $\tau$. The latter can be approximately related to the
logarithmic time derivative, $\beta$, of the nucleation rate,
$\tau\approx \beta^{-1}$. In adiabatic approximation one has
\be
\label{def_beta}
\frac{\beta}{H_*} = T_* \frac{d}{dT} 
\left. \left( \frac{S_3}{T} \right) \right|_{T_*},
\ee
where $H_*$ denotes the Hubble parameter at $T_*$. Hence, the typical
length scale is proportional to
\be
\left< R \right> \propto v_b \, \tau \approx \frac{v_b}{\beta}.
\ee
In contrast to Ref.~\cite{Apreda:2001us}, we state our results in
terms of the typical radius of the detonating bubbles $\left< R
\right>$ instead of using the approximate expression containing
$\beta$. It is not known what is the optimal definition for $\left< R
\right>$. Following Ref.~\cite{Nicolis:2003tg}, we use the maximum of
the bubble volume distribution, which should be closely related to the
typical scale for energy injection into the turbulent plasma. In this
case, one obtains
\be
\label{Rbeta_def}
\left< R \right> \approx 3\frac{v_b}{\beta}.
\ee
The relevant formulas can be found in the Appendix.

The energy density in GWs per logarithmic unit of frequency is usually
normalized to the critical density,
\be
\Omega(f)=\frac{1}{\rho_{\rm crit}}\frac{d \rho_{\rm GW}(f)}{d \ln f}. 
\ee
The GW contribution from colliding bubbles at the peak frequency, as
observed today, is then given by~\cite{Kamionkowski:1993fg}
\bea
h_0^2 \Omega_\textrm{det} &\simeq& 1.2 \times 10^{-7} \kappa^2 
\left< R \right>^2 H^2_* 
\left[ \frac{\alpha}{\alpha+1} \right]^2 
\frac{v_b}{0.24 + v_b^3} 
\left[ \frac{100}{g_*} \right]^{1/3}.
\eea
Assuming a detonation, i.e.~a strong phase transition with a
supersonic bubble wall, the wall velocity $v_b$ is approximately given
by \cite{Steinhardt:1981ct}
\be
v_b(\alpha) = \frac{1/\sqrt{3}+ \sqrt{\alpha^2 + 2 \alpha/3}}{1 + \alpha}.
\ee
This is a good approximation for very strong phase transitions.
Finally, the efficiency $\kappa$ is~\cite{Kamionkowski:1993fg}
\be
\label{eff_fac_kap}
\kappa(\alpha) \simeq \frac{1}{1 +0.715 \alpha}
\left[ 0.715 \alpha + \frac{4}{27} \sqrt{\frac{3\alpha}{2}} \right].
\ee
It measures the fraction of latent heat that is transformed into bulk
motion of the fluid.

The second source of GWs during the phase transition is
turbulence~\cite{Kamionkowski:1993fg,Kosowsky:2000rq,Dolgov:2002ra}.
When the bubbles collide, the plasma is stirred up, which creates a
cascade of eddies in the plasma. The relic GW density at the peak
frequency coming from turbulence, according to
Refs.~\cite{Dolgov:2002ra, Nicolis:2003tg}, is found to be
\bea \label{Dol}
h_0^2 \Omega_\textrm{turb} &\simeq& 5.6 \times 10^{-6} u_S^5  
 \left< R \right>^2 H^2_* 
\left[ \frac{100}{g_*} \right]^{1/3}.
\eea
As additional parameter, the characteristic turnover velocity $u_S$
of the eddies in the plasma enters, given by~\cite{Nicolis:2003tg}
\be
u_S \simeq \sqrt{\frac{\kappa \alpha}{ 4/3 + \kappa\alpha}}.
\ee
The expression (\ref{Dol}) relies on a certain modeling of time
correlations in the turbulence, which was introduced in
Ref.~\cite{Kosowsky:2001xp}. Similar results were obtained using
Kraichnan's ansatz for the time correlations
\cite{Gogoberidze:2007an}.

Recently, a different approach in calculating the contribution from
turbulence was presented in Ref.~\cite{Caprini:2006jb}. This work is
based on Richardson's model of turbulent motion which is based on
random velocities of the turbulent fluid. The authors of
Ref.~\cite{Caprini:2006jb} emphasize the fact that the GWs inherit the
momentum spectrum from the eddies rather than their frequency
spectrum. The resulting GW density at the peak frequency is
\bea
h_0^2 \Omega_{\rm Richardson} &\simeq& 6.7 \times 10^{-6} u_S^4 
  \left< R \right>^2 H^2_* 
\left[ \frac{100}{g_*} \right]^{1/3}.
\eea
Notice that if the results are stated in terms of $\left< R
\right>$, as done here, the numerical coefficients appearing in the GW
densities (and frequencies) of the two approaches to turbulence
basically agree. As mentioned before and detailed in
Ref.~\cite{Caprini:2006rd}, the main difference is based on the
question if the GWs inherit the momentum spectrum or the frequency
spectrum of the eddies resulting into an addition factor $u_S$ in the
GW density (and also in the peak frequency) and different scaling
behaviors.

On the other hand, comparing the numerical coefficients for turbulence
in terms of $\beta$ as it was presented in Ref.~\cite{Dolgov:2002ra}
and in the appendix of Ref.~\cite{Caprini:2006jb} one observes a
disagreement that can be traced back to the fact that different
expressions for relating the typical length scale $\left< R \right>$
to $\beta$ have beens used. In Ref.~\cite{Caprini:2006jb} the typical
bubble radius is given by $\left< R \right> \approx v_b / \beta$,
while in Ref.~\cite{Dolgov:2002ra} the relation $\left< R \right>
\approx 5\, v_b / \beta$ was used. To relate $\beta$ to $\left< R
\right>$ we use for both approaches the relation given
in~\eq~(\ref{Rbeta_def}) what is motivated in the appendix. Deciding
between these different approaches to turbulence is beyond the scope
of the current paper. We simply present results for both cases,
indicating the spread in the theoretical predictions.

For weaker phase transitions the contributions from bubble collisions are
larger than the ones from turbulence, but if the phase transition is
really strong, $u_S \sim v_b$, both contributions can be of comparable
magnitude or the latter can even dominate.

Since we assume the phase transition to lead to supersonic bubble wall
expansion and use the GW production formula at the frequency peak, the
given production rates of GWs should be understood as upper bounds.
If the wall velocity is subsonic, detonation does not take place and
the contribution of turbulence is strongly reduced, not only because
the velocity $v_b$ is small, but also because the efficiency factor $\kappa$
is much smaller than in \eq~(\ref{eff_fac_kap}), which as well assumes
supersonic bubble velocities.

To roughly determine the spectrum of the GWs, it is sufficient to know the peak
frequency and the scaling behavior. The peak frequencies are given 
by~\cite{Kamionkowski:1993fg, Dolgov:2002ra, Caprini:2006jb}
\bea
\label{peaks}
f_{\rm det} &\simeq& 1.6 \times 10^{-2} {\rm mHz} \, v_b
\, \frac{1}{\left< R \right> H_*} \, \frac{T_*}{100\,{\rm GeV}} 
\left[ \frac{g_*} {100}\right]^{1/6}, \\
f_{\rm turb} &\simeq& 1.7 \times 10^{-2} {\rm mHz} \, u_S
\, \frac{1}{\left< R \right> H_*}  \, \frac{T_*}{100\,{\rm GeV}} 
\left[ \frac{g_*} {100}\right]^{1/6}, \\ 
f_{\rm Richardson} &\simeq& 1.7 \times 10^{-2} {\rm mHz} 
\, \frac{1}{\left< R \right> H_*} \, \frac{T_*}{100\,{\rm GeV}}
\left[ \frac{g_*} {100}\right]^{1/6}. 
\eea
Notice that the peak frequency in Richardson's model differs from the
"usual" result for a turbulent plasma by a factor $u_S$ due to the
dispersion relation of the eddies. This shifts the peak frequency to
higher values.  This fact is relevant observationally, as the peak
frequencies for GWs from the electroweak phase transition are
typically at the lower end of the LISA window, where its sensitivity
is already reduced.

The GW spectrum from bubble collisions rises as $f^{2.8}$ below and
decreases as $f^{-1.8}$ above the peak~\cite{Kosowsky:1991ua}. The
corresponding scaling behavior in the case of turbulence is given by
$f^{2.0}$ and $f^{-3.5}$, respectively~\cite{Dolgov:2002ra}. Finally,
the scaling behavior in the analysis based on Richardson's model is
given by~\cite{Caprini:2006jb}
\be
h_0^2 \Omega_{\rm Richardson} \propto \left\{ {
\begin{matrix}
f^3,&      && f &<&      2 u_S f_\textrm{peak} \\
f,& 2 u_S f_\textrm{peak} &<&  f  &<& f_\textrm{peak} \\
f^{-8/3},& f_\textrm{peak}&<&  f  && \hskip 2cm .        \\
\end{matrix}}
\right. 
\ee

\section{The Standard Model with dimension-six operators \label{sec_phi6}}

Before turning to the nMSSM, let us discuss GW production in a simple
"toy model". It consists of the SM augmented by certain dimension-six
operators. These operators parametrize unknown physics at the cutoff
scale, typically at a few hundred GeV. They could originate from some
new strong dynamics, such as technicolor or weak scale gravity, or
from simply integrating out scalar fields.

The model contains a single Higgs doublet, $H$, whose potential is
stabilized by an $H^6$ interaction~\cite{Z93,GSW04,BFHS04}
\be
V(H)=-\frac{\mu^2}{2}H^2+\frac{\lambda}{4}H^4+\frac{1}{8M^2}H^6.
\ee
This potential has two free parameters, the suppression scale of the
dimension-six operator, $M$, and the quartic coupling, $\lambda$.  The
latter can be eliminated in terms of the physical Higgs mass,
$m_H$. Since the potential is stabilized by the $H^6$ term, $\lambda$
can be negative. In this case a barrier in the Higgs potential is
present at tree-level, which induces a strong first-order electroweak
phase transition.  Evaluating the one-loop thermal potential, it was
shown in Ref.~\cite{BFHS04} that the phase transition is strong enough
to avoid baryon number washout, i.e.~$\langle H\rangle_{T_c}/T_c>1.1$
\cite{M98}, if $M<850$ GeV and $m_H=115$ GeV. At the critical
temperature, $T_c$, the broken and symmetric phases are degenerate in
energy. Taking $M=500$ GeV, a strong phase transition is present for
$m_H<180 $ GeV.

Dimension-six operators also induce new sources of CP violation. In
addition to the ordinary Yukawa interaction of the top quark, $y_tH
t^cq_3$, one has an operator $(x_t/M^2)(H^{\dagger}H) \,H\,
t^cq_3$. Along the bubble wall these two contributions to the top
quark mass enter with varying weight. In this way a possible relative
phase between the two operators induces a varying complex phase in the
top quark mass such that tops and anti-tops behave differently in the
bubble background. Chiral charges are built up in front of the bubble
wall, which the sphalerons transform into a baryon asymmetry. A
semi-classical analysis of the these processes has been performed in
Refs.~\cite{BFHS04,FH06}. It was shown that the observed BAU can
indeed be produced this way. This can be achieved without generating
dangerously large electric dipole moments~\cite{HPR06}.

\begin{figure}[t]
\begin{center}
\includegraphics[width=0.95 \textwidth, clip]{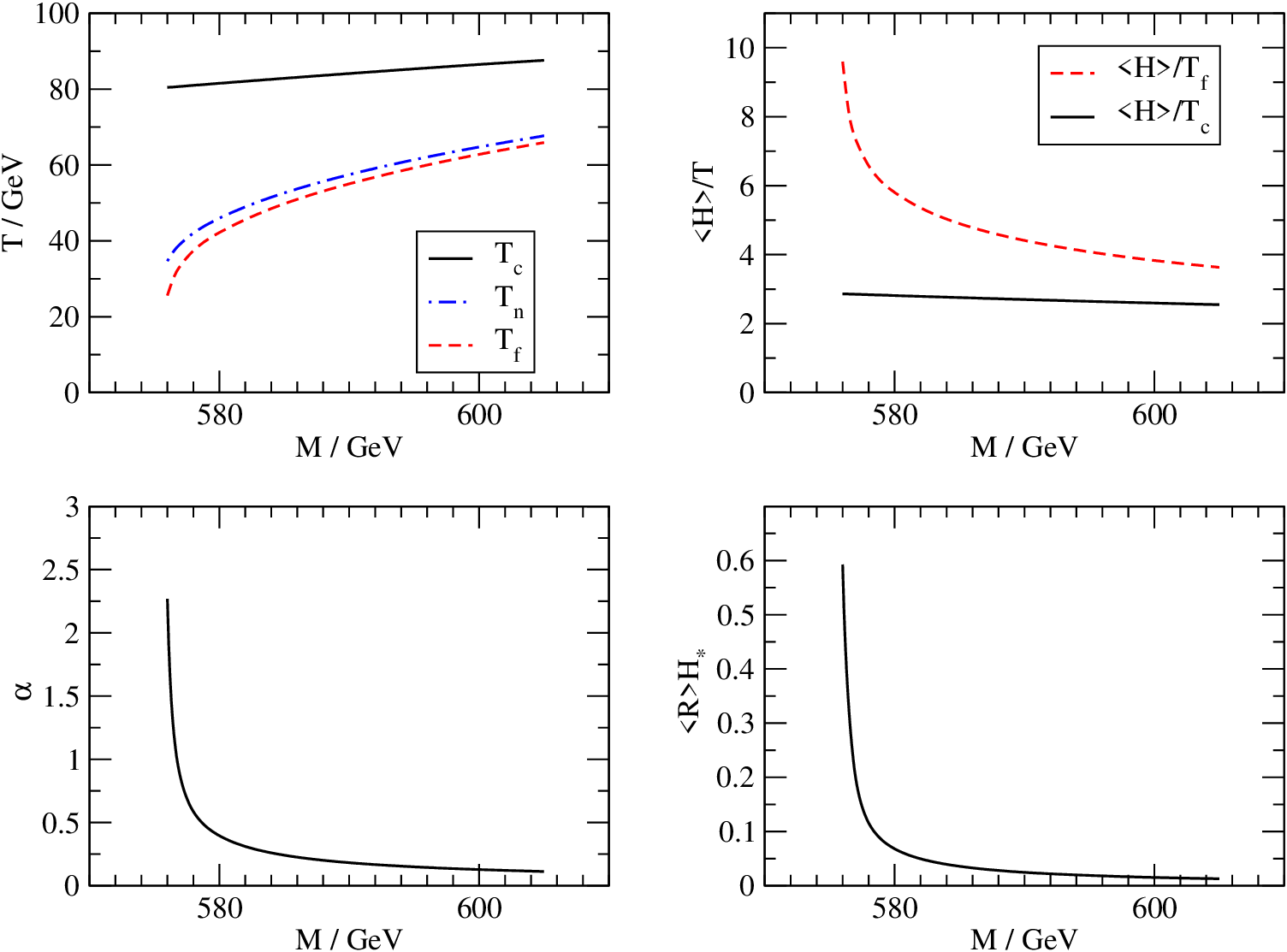}
\end{center}
\vskip -0.8cm
\caption{%
\label{fig_phi6_pt}
Different characteristics of the phase transition as functions of the
parameter $M$. The Higgs mass is chosen to be $m_h = 120$ GeV.}
\vskip 0.8cm
\end{figure}

To compute the spectrum of GWs one has to study the real time history
of the electroweak phase transition, in particular one must determine
the parameters $\alpha$ and $\langle R\rangle$. This requires
knowledge of the bubble nucleation rate, which depends on the energy
of the nucleating bubbles and varies with temperature. The bubble
configuration is a stationary solution to the Euclidean equations of
motion of the Higgs fields. The relevant formulas are collected in the
Appendix.  We use the one-loop thermal Higgs potential as described in
Ref.~\cite{BFHS04} but evaluate the full one-loop result
instead of using the high-temperature expansion. Since
there is only one Higgs field, the bubble solution can be obtained by
a simple over/under-shooting method.

We define the onset of the phase transition as the temperature, $T_n$,
where the first bubble per horizon volume nucleates, (\ref{eq_P}). As
the Universe cools down more bubbles nucleate, expand and finally
collide. During this stage one can neglect the expansion of the
Universe, which should only matter for extremely strong phase
transitions, $\beta \approx H_*$.  Most bubbles collide at the end of
the phase transition, which can be defined as the temperature, $T_f$
where a fraction of $1-1/e=0.632$ of the volume is converted into the
broken phase, (\ref{def_Tf}). Finally, we obtain $\langle R\rangle$
from the maximum of the bubble volume distribution at $T_f$, as
mentioned before.

\begin{figure}[t]
\begin{center}
\includegraphics[width=0.95 \textwidth, clip]{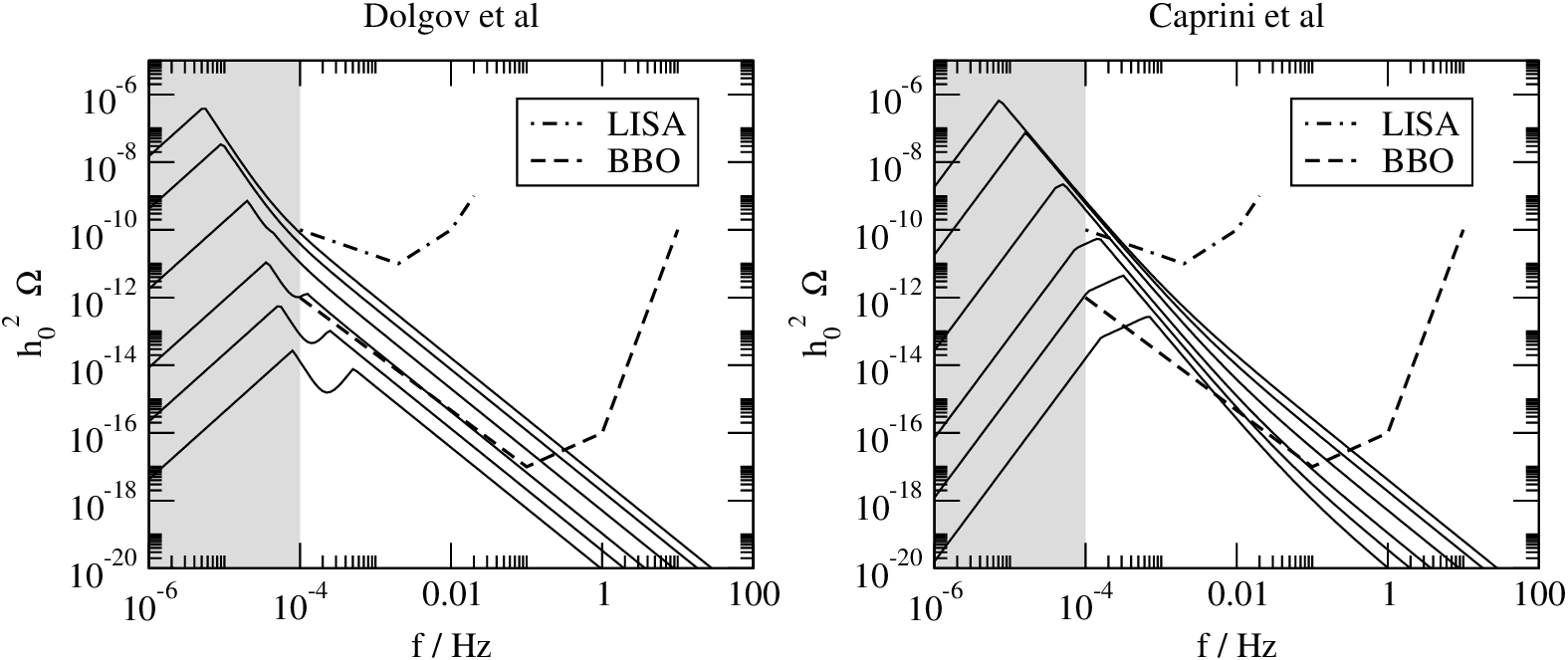}
\end{center}
\vskip -0.8cm
\caption{%
\label{fig_phi6_final}
The spectrum of the density of GWs for different values of the
parameter $M$. The used parameters are given in
Tab.~\ref{finaltab_phi6}. Smaller values of $M$ lead hereby to
a stronger phase transition and larger GW production. In the shaded
region, the sensitivity of LISA and BBO is expected to drop
considerably. In the left (right) panel, the turbulent
contribution to the GW spectrum from Ref.~\cite{Dolgov:2002ra}
(\cite{Caprini:2006jb}) has been used.}
\vskip 0.8cm
\end{figure}

\begin{table}
\begin{tabular}[b]{|c||c|c|c|c|}
\hline
\quad set \quad& 
\quad $M$ / GeV \quad &
\quad $\alpha$ \quad\quad &
\,\, $\left< R \right> H_*$ \quad &
\,\, $T_*$ / GeV \quad  \\
\hline
\hline
1 & 600.0 & 0.128 & 0.015 & 62.8 \\
\hline
2 & 588.0 & 0.201 & 0.028 & 53.1 \\
\hline
3 & 582.0 & 0.311 & 0.050 & 45.6 \\
\hline
4 & 578.0 & 0.586 & 0.116 & 37.4 \\
\hline
5 & 576.5 & 1.197 & 0.318 & 30.4 \\
\hline
6 & 576.0 & 2.268 & 0.592 & 25.6 \\
\hline
\end{tabular}
\caption{Sets of parameters used in Fig. \ref{fig_phi6_final}.
\label{finaltab_phi6}
}
\end{table}

In our numerical example we use the Higgs mass $m_H=120$. We have
checked that for larger Higgs masses the results are essentially the
same, only the relevant values of $M$ change. When $M$ decreases, the
the strength of the phase transition increases. At $M\sim 576$ GeV the
system becomes metastable and the Universe gets stuck in the symmetric
phase. As shown in Fig.~\ref{fig_phi6_pt} the phase transition is
strong enough to avoid washout after electroweak baryogenesis. While
$T_c$ and $\left< H \right> /T_c$ change only slowly in the presented
range of $M$, $T_f$ and $T_n$ drop rapidly. This signals that the
system is close to metastability.  At the same time $\alpha$ rises
from a value around $0.1$ to order unity and the typical bubble radius
$\left< R \right>$ increases dramatically. In
Fig.~\ref{fig_phi6_final} the GW spectrum for several choices of $M$
is shown, as given in Tab.~\ref{finaltab_phi6}, together with the LISA
and BBO sensitivities. In gray the region of $f<10^{-4}$ Hz is
indicated, where observational sensitivities drop considerably.
In the two panels, the two different predictions for GW
production from turbulence as given in Refs.~\cite{Dolgov:2002ra} and
\cite{Caprini:2006jb} have been used. The plot demonstrates that as
the phase transition becomes stronger and the amplitude of the GW
signal becomes larger, the peak frequency moves to lower values. This
is directly related to the smaller values $T_f$ and larger values of
$\langle R\rangle H_*$ in Fig.~\ref{fig_phi6_pt}.  As a result, even
in the case of an extremely strong transition, $M\sim 576$~GeV, it
will be difficult for LISA to detect a GW signal. This would require
sensitivity at lower frequencies. Things are different for BBO. It
would be able to detect GWs if $M<585$~GeV, which corresponds to a few
percent tuning in $M$.  In this case also the subtraction of
backgrounds, in particular from white dwarf binaries, has to be under
control (see
\cite{Grojean:2006bp}, and references therein).  The main difference
between the two models of turbulence is that in the case of the
Richardson model the two-peak structure in the GW spectrum disappears,
as the turbulence contribution is shifted to higher frequencies. If
present and observed, the double-peak structure could help to identify
a phase transition as source of the GW signal. Comparing with the
general analysis in Ref.~\cite{Grojean:2006bp} we arrive at a more
pessimistic picture because large values of $\alpha$ mean small
$\langle R\rangle$ and a low peak frequency. The decrease in
temperature for very strong phase transitions additionally enhances
this effect.

\section{The nMSSM\label{sec_model}}

While the electroweak phase transition in the MSSM is too weak to
produce observable GWs~\cite{Apreda:2001us}, the strength of the phase
transition increases considerably if the MSSM is equipped with an
additional gauge singlet \cite{NMSSM}. There are various possibilities
to implement this idea. A phenomenologically attractive candidate is
the nMSSM \cite{nMSSM} with the superpotential
\be
W_{\rm nMSSM} = \lambda \hat S \hat H_1 \cdot \hat H_2
- \frac{m^2_{12}}{\lambda} \hat S
+ W_{\rm MSSM}.
\label{W_nMSSM}
\ee
In contrast to the models studied in Ref.~\cite{NMSSM}, it solves the
$\mu$-problem of the MSSM without destabilizing the electroweak
hierarchy or the generation of domain walls at the electroweak phase
transition. However, the discrete R-symmetry necessary to accomplish
this task, forbids the singlet self-coupling and leads to a quite
constrained Higgs and neutralino/chargino
phenomenology~\cite{Menon:2004wv,Huber:2006wf}. In particular,
the nMSSM does not contain a cubic singlet term in the superpotential,
unlike the NMSSM. Hence, the singlino obtains its mass only by mixing
with the Higgsinos. It is typically very light, and can be a dark
matter candidate \cite{Menon:2004wv}. An early version of a singlet
model without a self-coupling can be found in
Ref.~\cite{Fayet:1974pd}.

The tree-level Higgs potential of the nMSSM including soft SUSY
breaking terms reads
\be 
V=V_F + V_D + V_{\rm soft}, 
\ee
where
\bea
\label{scalar_pot}
V_F &=&  \lambda^2 |H_1 \cdot H_2 |^2
       + \lambda^2 |S|^2 (H_1^\dagger H_1 + H_2^\dagger H_2) 
       -  (m_{12}^2 H_1 \cdot H_2 + \, h.c. ), \\
V_D &=& \frac{g^2+{g'}^2}{8} (H_2^\dagger H_2 - H_1^\dagger H_1)^2 +
	\frac{g^2}{2} |H_1^\dagger H_2|^2, \\
V_{\rm soft} &=& m_1^2 H_1^\dagger H_1 + m_2^2 H_2^\dagger H_2 
  	+  m_s^2 |S|^2 +  (t_s S + \, h.c.) + (a_\lambda S H_1 \cdot H_2 + \, h.c.).
\eea
Additionally, we take into account the Coleman--Weinberg one-loop terms
\bea
\Delta V = \frac{1}{16 \pi^2} \left[ \sum_b g_b \, h(m^2_b)
- \sum_f g_f \, h(m^2_f)  \right], \label{C-W-log}
\eea
where the two sums run over bosons and fermions with the
degrees of freedom $g_b$ and $g_f$ respectively, and
\bea
h(m^2) = \frac{m^4}{4} \left[ \ln \left( \frac{m^2}{Q^2} \right) - \frac32 \right].
\eea
At finite temperature, the effective potential is modified by the
following thermal one-loop contributions:
\bea
\Delta V^T = \frac{T^4}{2 \pi^2} \left[ \sum_b  g_b J_+ (m_b^2 / T^2) -
\sum_f  g_f J_- (m_f^2 / T^2) \right],
\eea
with
\bea
J_\pm(y^2) = \int_0^\infty dx \, x^2 \, \log( 1\mp \exp(- \sqrt{x^2 + y^2}) ).
\eea
For both one-loop contributions we use the following numbers of
degrees of freedom
\be
g_W = 6, \quad g_Z = 3, \quad g_t = 12, \quad g_{\tilde t_1} = g_{\tilde t_2} = 6,
\ee
and the renormalization scale is chosen to be $Q=150$ GeV. As
in Ref.~\cite{Huber:2006wf}, the stop masses are chosen as
\be
m_U = m_Q = 500 \textrm{ GeV},
\ee
such that the strength of the phase transition originates mostly from
the (tree level) singlet sector. As most SUSY particles are assumed to
be heavy, we approximate the number of effective degrees of freedom by
the SM value, $g_* = 107.75$.

At zero temperature, both Higgs fields and the singlet field acquire a
vev and the Higgs vevs are constrained by
\be
\phi(T=0) = v \simeq 174 \textrm{ GeV}, \quad
\phi^2(T) = |\left< H^0_1 \right>_T|^2 + |\left< H^0_2 \right>_T|^2.
\ee
Additionally, since one of the chargino masses is approximately given by 
\be
m_{\chi^\pm} \approx \lambda \left< S \right> \gtrsim 114 \, {\rm GeV},
\ee
the parameter space is restricted to regions with a sufficiently large
singlet vev. The lower bounds on the masses of the physical Higgs
states further restrict the parameter space, such that the nMSSM
constitutes a much stronger constrained model than models with an
explicit Higgsino/chargino mass ($\mu$-term) or a singlet
self-coupling.  These bounds are easier to satisfy for larger values
of $\lambda$, which therefore prefers to be close (even beyond) the
Landau pole value.

\section{Numerical Analysis\label{sec_num}}

Our numerical analysis follows the approach already used in
Ref.~\cite{Huber:2006wf} to determine the BAU in the nMSSM. The
relevant eight free parameters of the nMSSM are chosen randomly.  The
generated parameter sets are then confronted with several constraints
on the mass spectrum and the Z-width (for details see
Ref.~\cite{Huber:2006wf}). In addition, we restrict ourselves to cases
which promise a rather strong phase transition, $\phi(T_c)/T_c
\gtrsim 1$, by inspection of the potential. If the parameter set
passes these constraints, the properties of the phase transition and
the GW relic density are determined.

Compared to the former work on the GW production in the
NMSSM~\cite{Apreda:2001us}, the analysis is improved in the following
points. First, we use up-to-date bounds on the particle spectrum,
which severely constrains the parameter space. Second, the bubble
configurations of the six scalar fields are determined exactly, while
in Ref.~\cite{Apreda:2001us} this problem was reduced effectively to a
one-dimensional one, which overestimates the strength of the phase
transition and hence the GW production.  Third, the parameters
relevant for baryogenesis are determined in the specific cases and
hence a direct correlation between the BAU and the magnitude of the
GWs can be inferred.

Because of the linear and tri-linear terms of the singlet in the
potential~(\ref{scalar_pot}), a strongly first-order phase transition
in the nMSSM is possible due to tree-level
dynamics~\cite{Hugonie:2003yu, Ham:2004nv, Huber:2006wf,
Menon:2004wv}. The phase transition in this case is described by six
fields: The two vevs of the neutral Higgs fields $\left<
H_1^0\right>$, $\left< H_2^0\right>$, the vev of the singlet $\left< S
\right>$ and their complex phases; however, the relative phase between
the Higgs vevs does not enter the potential and can be removed using
its equation of motion. In this work, the profiles of the vevs during
the phase transition are calculated numerically using the method
presented in Ref.~\cite{Konstandin:2006nd} that is briefly described
in Appendix~\ref{sec_PT}.

\begin{figure}[t]
\begin{center}
\includegraphics[width=0.95 \textwidth, clip]{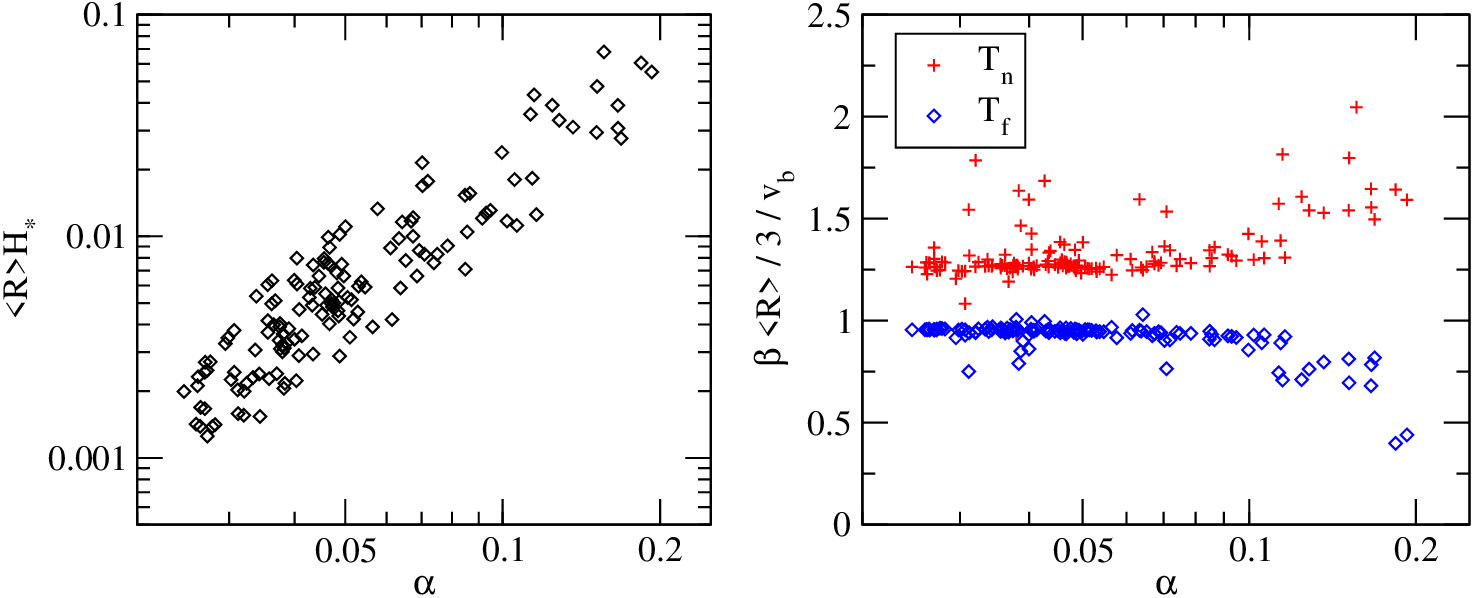}
\end{center}
\vskip -0.8cm
\caption{%
\label{fig_ab}
The left panel shows the correlation between the parameters $\alpha$
and the typical size of the bubbles at the end of the phase
transition, $\left< R \right>$. The right panel compares the typical
bubble size with $\beta$ at first nucleation, $T=T_n$, and the end of
the phase transition, $T=T_f$.  }
\vskip 0.8cm
\end{figure}
The left panel of fig.~\ref{fig_ab} shows the correlation between the
parameters $\alpha$, the available energy, and $\left< R \right>H_*$ for
approximately 150 parameter sets. We find the same behavior as in the
SM with dimension-six operators. The right panel compares the typical
bubble size with $\beta$ at first nucleation, $T=T_n$, and the end of
the phase transition, $T=T_f$. Note that $\beta(T_n)$ usually
underestimates the bubble radius, while $\beta(T_f)$ overestimates
it. For strong phase transitions, $\beta$ varies considerably between
$T_n$ and $T_f$. This makes our more careful determination of $\left<
R \right>$ necessary. The same applies to the SM with dimension-six
operators.

Since there is a relatively strong correlation between $\alpha$ and
the bubble radius, we correlate in the following only the parameter
$\alpha$ with the parameters that enter the determination of the BAU
and the production of GWs.
\begin{figure}[t]
\begin{center}
\includegraphics[width=0.95 \textwidth, clip]{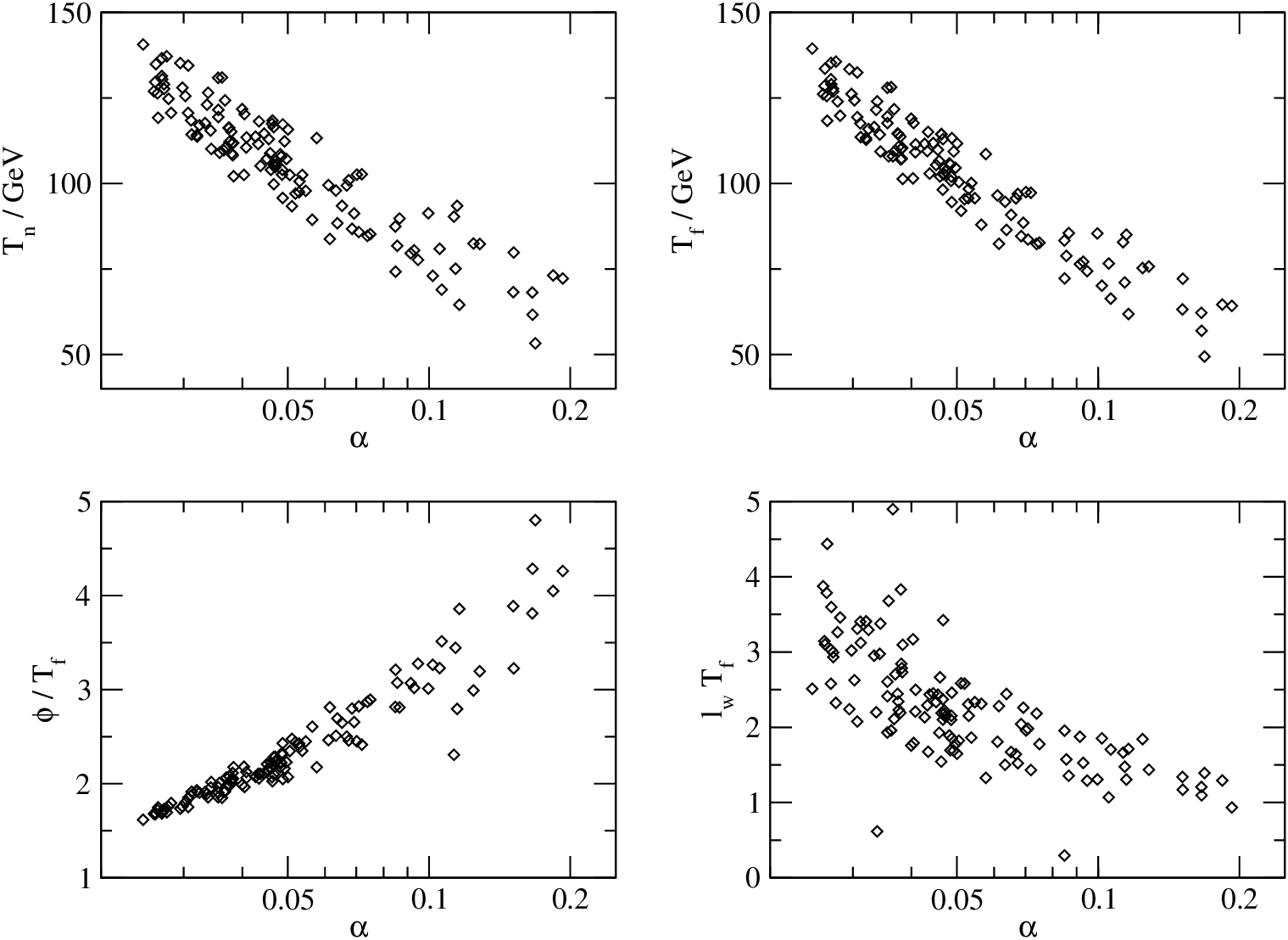}
\end{center}
\vskip -0.8cm
\caption{%
\label{fig_alphPT}
The correlation between the parameter $\alpha$ and several properties
of the phase transition: The temperature at first nucleation $T_n$,
the temperature at the end of the phase transition $T_f$, the
thickness of the bubble wall $l_w$ and the ratio between the Higgs vev
and the temperature $\phi(T_f)/T_f$.}
\vskip 0.8cm
\end{figure}

Figure~\ref{fig_alphPT} shows several characteristic properties of the
phase transition. In all parameter sets under consideration, baryon
number washout after the phase transition is sufficiently
suppressed~\cite{M98}, $\phi(T_f)/T_f \gtrsim 1.1$. As concluded in
Ref.~\cite{Huber:2006wf}, electroweak baryogenesis is a generic
feature for this class of models. However, this statement is based on
the assumption that the wall velocity is subsonic, such that diffusion
is operative.
From the parameters that enter  the generation of the BAU in
\eq~(\ref{eta10}), three are expected to have a strong dependence on the
strength of the phase transition: The wall thickness $l_w$ and the
temperatures $T_n$ and $T_f$. These quantities are also plotted
against the parameter $\alpha$ in Fig.~\ref{fig_alphPT}. Both, the
wall thickness and the temperatures, decrease with increasing strength
of the phase transition.

\begin{figure}[t]
\begin{center}
\includegraphics[width=0.95 \textwidth, clip]{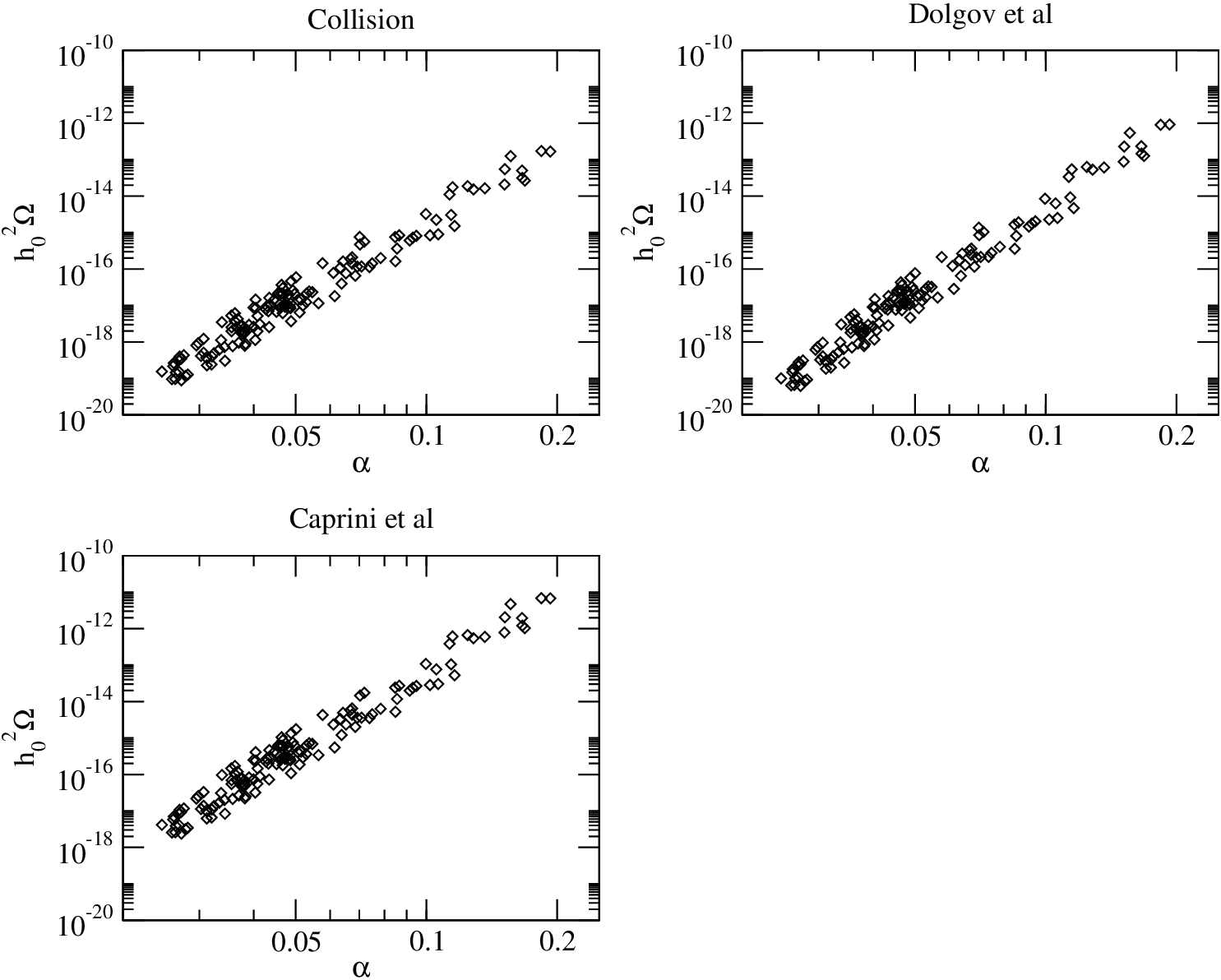}
\end{center}
\vskip -0.8cm
\caption{%
\label{fig_bau}
The magnitude of the produced gravitational waves for our set of
models. The panels show the GWs from collisions and from turbulence
following the two different approaches under consideration.}
\vskip 0.8cm
\end{figure}

The magnitude of the produced gravitational waves is displayed in
Fig.~\ref{fig_bau}, separately for the contribution from collisions
and turbulence.
For our parameter sets, the strength of the phase transition never
exceeds $\alpha=0.2$ and the density of the produced GWs is too low to
be observed by the LISA experiment that has a sensitivity of
$h_0^2\Omega
\sim 10^{-11}$.
\begin{figure}[t]
\begin{center}
\includegraphics[width=0.95 \textwidth, clip]{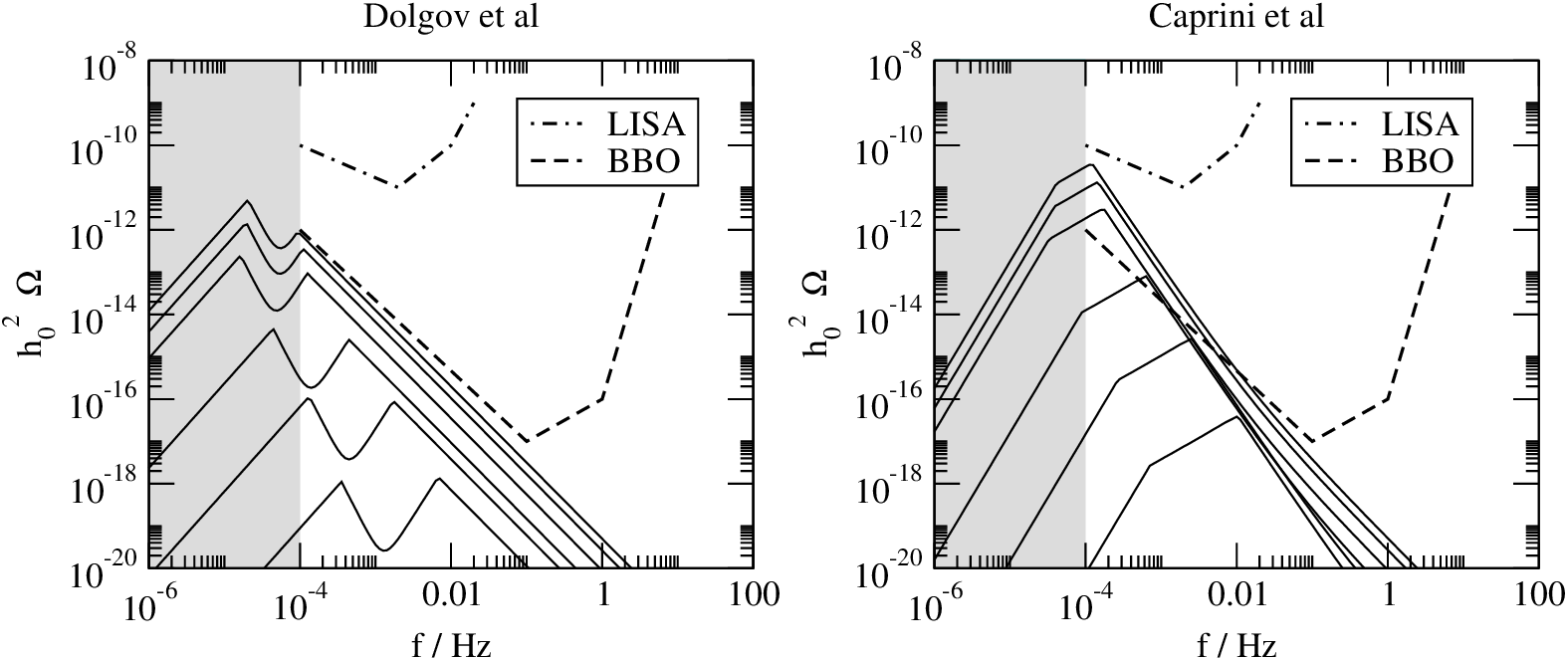}
\end{center}
\vskip -0.8cm
\caption{%
\label{fig_final}
The spectrum of the density of GWs for different nMSSM models. The
used parameters are given in Tab.~\ref{finaltab}. In the
shaded region, the sensitivity of LISA and BBO is expected to drop
considerably. In the left (right) panel, the turbulent
contribution to the GW spectrum from Ref.~\cite{Dolgov:2002ra}
(\cite{Caprini:2006jb}) has been used.}
\vskip 0.8cm
\end{figure}
On the other hand, the BBO experiment has a maximal sensitivity of
$h_0^2\Omega \sim 10^{-17}$ which is several orders below the maximal  
GW density we found, $h_0^2\Omega \sim 10^{-11}$.

However, even the prospects for the BBO experiment to observe GWs from
a strongly first-order phase transition in the nMSSM are very limited
as seen in Fig. \ref{fig_final}.  The plot shows six spectra
corresponding to the parameters given in Table \ref{finaltab}. Even
for relatively strong phase transitions, the spectrum of GWs would
escape detection, since a stronger phase transition proceeds at lower
temperatures and leads to larger bubbles, as can be seen in
Fig. \ref{fig_alphPT}. This effect lowers the peak frequencies
according to \eqs~(\ref{peaks}) to regions where the sensitivity of
BBO is strongly reduced. This is the same phenomenon we already
observed in the SM with dimension-six operators.

\begin{table}
\begin{tabular}[b]{|c||c|c|c|}
\hline
\quad set \quad& 
\quad $\alpha$ \quad\quad &
\quad  $\left< R \right> H_*$ \quad &
\quad $T_*$ / GeV \quad  \\
\hline
\hline
1 & 0.03 & 0.003 & 130 \\
\hline
2 & 0.05 & 0.01 & 110 \\
\hline
3 & 0.07 & 0.03 & 85 \\
\hline
4 & 0.1 & 0.1 & 80 \\
\hline
5 & 0.15 & 0.1 & 70 \\
\hline
6 & 0.2 & 0.1 & 60 \\
\hline
\end{tabular}
\caption{Sets of parameters used in Fig. \ref{fig_final}.\label{finaltab}
}
\end{table}

This leaves the question why the nMSSM does not seem to have a region
in parameter space where $\alpha$ is very large, as observed in the
NMSSM in Ref.~\cite{Apreda:2001us}. One might think that the nMSSM is
much more constrained than the NMSSM and that Higgs and chargino
masses in accordance with current LEP bounds strongly restrict the
parameter space of viable models. Nevertheless, the nMSSM admits very
strong phase transitions induced by tree-level terms, even leading to
metastability.  So the crucial difference seems to be that we
determine the tunnel action exactly, while in the
work~\cite{Apreda:2001us} the problem was reduced to one dimension,
which can considerably overestimate the strength of the phase
transition.

According to \eq~(\ref{alpha_def}), large $\alpha$ requires either
large amounts of latent heat or small nucleation temperatures. The
latent heat cannot be made arbitrarily large. For example, very
small/large values for $a_\lambda$ or $t_s$ can lead to an unbounded
effective one-loop potential or are not compatible with the mass
constraints (one explicit example is given in
Ref.~\cite{Huber:2006wf}). Thus, the upper bound of the latent heat is
expected to be limited approximately by the electroweak scale ($v
\simeq 174$ GeV) and stronger phase transitions require smaller
temperatures
\be
\label{scale_1}
\alpha \lesssim {\cal O}\left( \frac{30}{\pi^2 g_*}\frac{v^4}{T_*^4} \right).
\ee
On the other hand, the nucleation temperature $T_*$ cannot be many
orders smaller than the temperature $T_c$, where the two minima of the
effective potential are degenerate. This can be seen in the following
way. Assume that the Higgs vev is close to its zero temperature value,
$\phi(T)\approx v$, and that the potential is of the form
\be
\label{delV_approx}
\Delta V(T) = \Delta V(0) + \tilde g^2 \, T^2\, v^2 
= \tilde g^2 \, (T^2-T^2_{c})\, v^2,
\ee
with some constant $\tilde g$ that mostly depends on the particle
content of the theory. Since the three-dimensional action $S_3$ scales
as $\Delta V^{-2}$ for small $\Delta V$, the condition in
\eq~(\ref{Tn_approx}) turns into
\be
\left. \frac{S_3(T)}{T} \right|_{T_*} = \frac{S_3(0)}{T_*} 
\frac{T_{c}^4}{(T_{c}^2-T_*^2)^2}  \approx 140.
\ee
This implies $T_* > \frac15 T_{c}$, since the critical temperature has
to be larger than the minimum of this function.  The numerical results
in the nMSSM show that the minimum is rather given by $T_* > c\,
T_{c}$ with the constant $c\approx 0.5$. This is mainly due to the
fact that the three-dimensional action only scales as $\Delta V^{-2}$
for very small $\Delta V$.  The closer $T_*$ is to the minimum, the
larger is $\alpha$ and the smaller is $\beta$.  This in turn leads to
the fact that small critical temperatures require small latent heat,
since one obtains, using \eq~(\ref{delV_approx}),
\be
|\Delta V(0)| < \epsilon_* = \tilde g^2 \, (T^2_{c}+T^2_*)\, v^2 
< (1+c^{-2}) \, \tilde g^2 \, T^2_*\, v^2.
\ee
Thus, the parameter $\alpha$ scales in this limit as
\be
\label{scale_2}
\alpha \propto \frac{\Delta V}{T_*^4} 
\propto \frac{1}{\Delta V} \propto \frac{1}{ T_*^2}.
\ee

Moreover, small latent heat $\epsilon_*$ and thus small $|\Delta V(0)|$
tend to increase the three-dimensional tunnel action 
\be
140 \ T_* \approx S_3(T_*)  \propto \Delta V(T_*)^{-2}  
>  \Delta V(0)^{-2}, 
\ee    
such that the condition in \eq~(\ref{Tn_approx}) implies that large
$\alpha$ not only requires a small $|\Delta V(0)|$, but also a very
small potential well between the two minima. Our numerical analysis
indicates that this seems not to be compatible with the mass
constraints on the charginos and Higgs bosons in the nMSSM, what
limits the strength of the phase transition. Nevertheless, with
further tuning a somewhat larger value for $\alpha$ might be possible.

\section{Electroweak Baryogenesis}

While the plasma is driven out of equilibrium by the expansion of the
Higgs bubbles during the phase transition, CP violation is provided by
the complex coefficients in the potential of \eq~(\ref{scalar_pot}).
After redefinition of the fields, all sources of CP violation can be
attributed solely to the parameter $t_s$ by a complex phase. (We do
not consider a CP-violating phase in the Wino mass, as was done in
Ref.~\cite{Menon:2004wv}.)  As mentioned before, the nMSSM does not
contain an explicit Higgsino mass $\mu$, but only an induced Higgsino
mass from the singlet-Higgs coupling according to
\be
\mu = \lambda \left< S \right>.
\ee
The change in the complex phase of the singlet during the phase
transition gives rise to a CP-violating current of charginos in the
bubble wall. Via Yukawa interactions, this CP violation is communicated to
the tops that bias the sphaleron process. This mechanism is called
chargino mediated electroweak baryogenesis and was analyzed in the
MSSM in
Refs.~\cite{Cline:2000nw,Carena:2002ss,Carena:2000id,EWBG_series}. The
formalism we use to determine the BAU was developed in the series of
papers~\cite{EWBG_series} and recently applied to the
nMSSM~\cite{Huber:2006wf}.

In the case of a large Wino mass parameter, $M_2 \gtrsim 1$~TeV, the
change in the complex phase, $\Delta{\rm Arg}(\mu)$, during the phase
transition is the main source of baryon number generation and a good
estimate of the predicted $\eta_{10}$ is given by
\be
\eta_{10} \approx \, c(T_*) \,\,\frac{\Delta {\rm Arg}(\mu)}{\pi} 
\frac{1}{ \, l_w \, T_* } \,
\left( \frac{\mu_0}{\tau T_*} \right)^{\frac32} \, \frac{\Delta\mu}{\tau T_*} \,
\exp(-\mu_0 /\tau T_* ), \label{eta10}
\ee
where $\mu_0$ denotes the $\mu$ parameter in the symmetric phase,
$\Delta\mu$ the change in $\mu$ during the phase transition and $l_w$
the thickness of the bubble wall. The two coefficients are $c(T_*)
\approx 1.6 \,T_*/$GeV and $\tau \approx 0.78$.  
\begin{figure}[t]
\begin{center}
\includegraphics[width=0.7 \textwidth, clip]{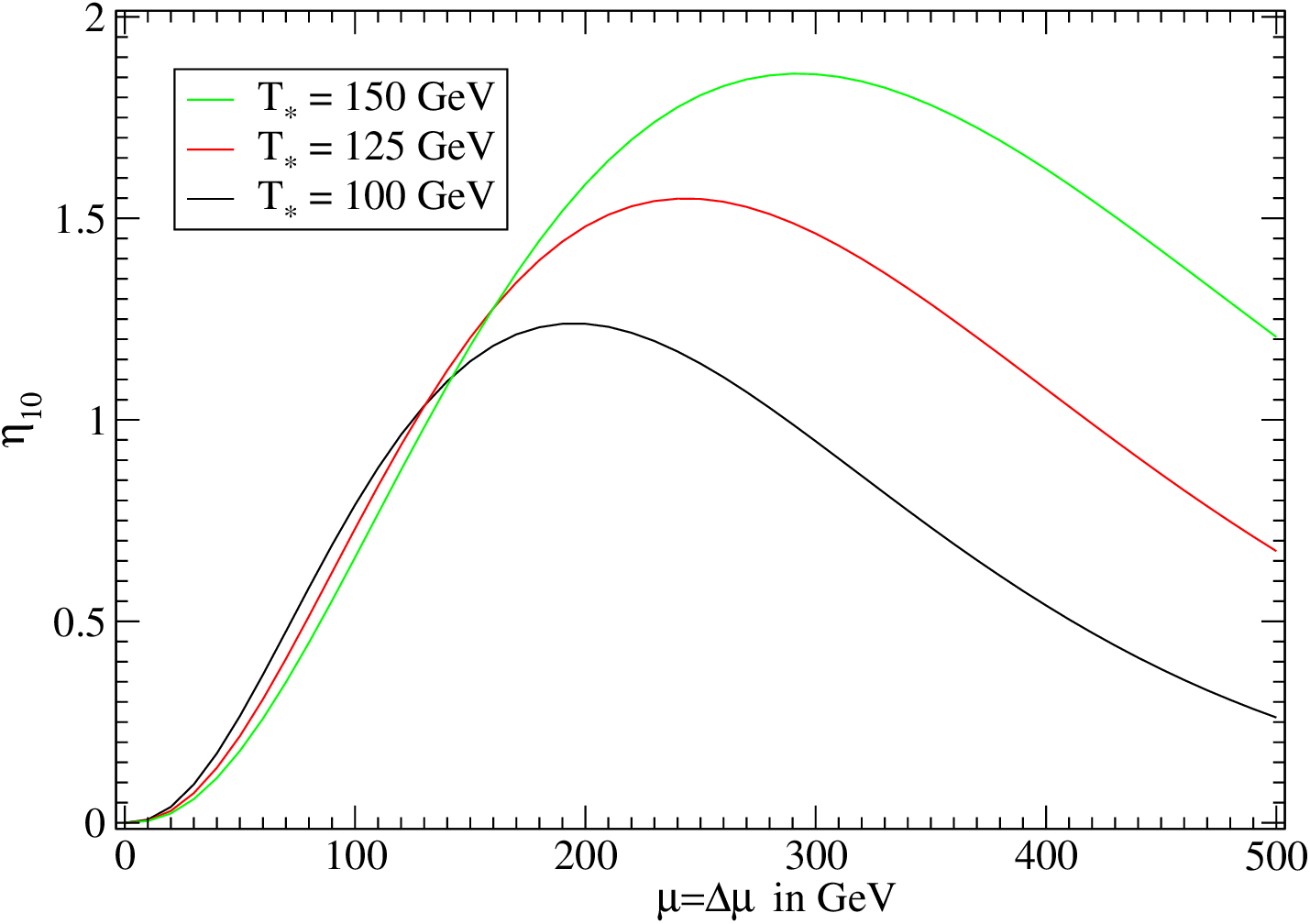}
\end{center}
\vskip -0.8cm
\caption{%
\label{fig_test}
The baryon asymmetry $\eta_{10}$ for different critical temperatures
$T_*$, ${\Delta {\rm Arg}(\mu)}=\pi/10$ and $l_w \, T_* = 10$.}
\vskip 0.8cm
\end{figure}
The function $\eta_{10}$ is plotted in Fig.~\ref{fig_test} for
$\mu=\Delta\mu$ and several values of $T_*$ and shows very good
agreement with the full numerical results of the diffusion equations
given in Ref.~\cite{Huber:2006wf}. This has to be compared with
the observed baryon asymmetry that, normalized to the entropy density,
is given by~\cite{Spergel:2003cb}
\be
\eta_{10} = \frac{n_B - n_{\bar B}}{s} 10^{10} = 0.87 \pm 0.03.
\ee
We see that under these assumptions, electroweak baryogenesis is a
quite generic feature of the nMSSM. However, these results are not
directly applicable to the present case, since they have been obtained
using a wall velocity smaller than the speed of sound. In this regime,
the produced BAU does not strongly depend on the precise velocity of
the wall. In the context of baryogenesis and for not too strong phase
transitions, subsonic bubble velocities are motivated by calculations
of the friction effects acting on the bubble wall~\cite{Moore:1995si,
Moore:2000wx, John:2000zq}.

For the case at hand, the phase transition is usually strong enough to
make friction negligible such that the phase transition indeed
proceeds by detonation and with supersonic wall velocities, as can be
estimated in the following way. According to Ref.~\cite{Moore:2000wx},
the main friction on the wall constitutes the interaction of the W
bosons and is given by
\be
\eta \sim \frac{3 m_D^2 T_*}{16\pi l_w} \log(m_W l_w),
\ee
where $m_W$ and $m_D$ denote the mass and Debye mass of the W bosons
in the plasma. On the other hand, the expansion of the bubbles is
driven by the latent heat. The resulting pressure difference acting on
the wall can be estimated to be
\be
\Delta p \sim \frac13 \epsilon_* = \alpha \frac{ \pi^2 g_* T_*^4}{90}.
\ee
Thus, in the present case friction from the W bosons can be neglected,
$\eta \ll \Delta p$, as long as $\alpha \gg 0.01$ and 
the phase transition proceeds by detonation. 

As has been explained earlier, for an observable production of GWs,
wall velocities beyond the speed of sound are necessary. This is
problematic for standard electroweak baryogenesis, since in this
regime diffusion processes are strongly suppressed. However, even if
the analysis based on diffusion is not valid, other mechanisms, as
e.g. quantum mechanical reflection at the
wall~\cite{Shaposhnikov:1991cu} might be responsible for the
production of the baryon asymmetry. The relevant parameters that enter
a quantitative analysis of the produced BAU are in fact similar. For
example, in the case of the nMSSM with a very strong phase transition,
one would expect that the produced baryon asymmetry is small since the
low temperature leads to Boltzmann suppressed chargino densities. This
problem might be avoided in scenarios where the CP violation is
supplied by SM particles that are massless in the symmetric phase, as
e.g.~in the model with dimension-six operators discussed in
Ref.~\cite{BFHS04,FH06}. Nevertheless, a strongly first-order phase
transition with supersonic bubble velocities seems to disfavor baryon
production at the electroweak scale.

\section{Conclusions\label{sec_con}}

In conclusion, we found that the density of GWs produced during a
strongly first-order electroweak phase transition is in a large
portion of parameter space too small to be detectable by LISA or even
the BBO experiment, both for the nMSSM and the Standard Model with
dimension-six operators. This is partially because stronger phase
transitions proceed at lower temperatures and create larger bubbles,
which shifts the peak of the spectrum of the GW density to lower
frequencies. This correlation was not taken into account in the model
independent analysis of Ref.~\cite{Grojean:2006bp}, which therefore
led to more optimistic conclusions.  For the nMSSM our results are
best summarized by Fig.~\ref{fig_final} which displays the spectrum of
GW densities for several parameter sets, and for the Standard Model
with dimension-six operators by Fig.~\ref{fig_phi6_final}. Moreover,
we observe that very strong phase transitions, $\alpha>0.2$, in the
nMSSM are not compatible with constraints on the particle spectrum, or
at least require a significant amount of tuning. The situation in the
Standard Model with dimension-six operators is slightly more
promising. Besides, it is easier to understand the effects of
parameter tuning in this case. Generating observable GWs at BBO
requires a few percent tuning in the suppression scale of the
dimension-six operator.

Technically, our main improvement was
the full numerical computation of the bubble configurations, which leads to
more realistic predictions on the properties of the phase transition that
enter the GW computation. The simple approximation to bubble profiles in
Ref.~\cite{Apreda:2001us} seems to considerably overestimate the strength of the 
phase transition and in turn the GW signal.

On electroweak scales, a strongly first-order phase transition is a
necessary prerequisite for electroweak baryogenesis as well as for
sizable GW production. However, there is no positive correlation
between GW observations and electroweak baryogenesis. In the nMSSM,
for all parameter sets under consideration, washout is suppressed
after the phase transition and electroweak baryogenesis might be
possible. Besides, GW production is marginal for many parameter sets,
such that electroweak baryogenesis cannot be excluded, even if GWs are
not observed down to a sensitivity of $h_0^2 \Omega=10^{-20}$. On the
other hand, observing GWs in the near future does not directly support
electroweak baryogenesis. In fact, the opposite might be true. Since
significant GW production by collisions and turbulence requires
supersonic bubble wall expansion, diffusion is suppressed which
reduces the produced BAU in mechanisms based on transport. Due to
rather thin walls, other mechanisms, as e.g. CP-violating reflection,
might be effective in this regime, but reliable predictions in this
scenario are missing. Additionally, in the case of the nMSSM,
observable GW production is only possible (if at all) for small
critical temperatures, what further reduces the prospects of chargino
mediated baryogenesis due to Boltzmann suppressed chargino densities,
even if they are not based on transport.

Hence, the observation of GWs that can be attributed to the
electroweak phase transition, could not only partially shadow a
smoking gun signal from inflation, but also question (at least the
standard picture of) electroweak baryogenesis.

It would be interesting to study baryogenesis with supersonic bubbles
in more detail. After all, supersonic walls could appear in a larger
part of the nMSSM parameter space. Finally, one should keep in mind
that especially the turbulence contribution to the GW spectrum is
still under debate. The available computations are based on
non-relativistic models of turbulence. A numerical simulation of
relativistic turbulence would be desirable to obtain more reliable
predictions. Also the high frequency behavior of the contributions
from bubble collisions is debatable. This was recently discussed using
an analytic approach in Ref.~\cite{Caprini:2007xq} and will be subject
of a forthcoming publication~\cite{HKneu}.

\section*{Acknowledgments}

 T.K. is supported by the Swedish Research Council
 (Vetenskapsr{\aa}det), Contract No.~621-2001-1611.

\appendix

\section{The Phase Transition\label{sec_PT}}

If the two minima of the effective potential are separated by a
potential well, thermal and quantum fluctuations lead to small regions
in space that acquire a finite Higgs vev close to the global minimum
of the effective potential. If this region reaches a critical size, it
is advantageous to expand this region, since the gain in energy by
increasing the size of the bubble dominates over the increase in
energy in the surface of the bubble. The phase transition proceeds
then by nucleation and expansion of bubbles consisting of regions with
non-vanishing Higgs vev.

In the semi-classical theory of
tunneling~\cite{Coleman:1977py,Linde:1980tt}, the tunnel probability
depends on the action of the so-called bounce solution. This
configuration fulfills the boundary conditions ($\varphi_-$ denotes
the symmetric minimum of the potential $V(\varphi,T)$)
\be
\partial_\varrho \varphi(0)=0, \quad  
\lim_{\varrho\to\infty} \varphi(\varrho)= \varphi_-
\ee
and the Euclidean equation of motion 
\begin{equation}
\label{EOM:Eu}
\frac{\partial^2\varphi}{\partial \varrho^2}+\frac{\gamma}{\varrho}
\frac{\partial\varphi}{\partial \varrho} = V^\prime(\varphi)\,,
\end{equation}
and $\gamma=2(3)$ corresponds to tunneling at finite temperature (in
vacuum). At finite temperature, the bubble nucleation rate is then
given by
\be
\Gamma = A \, T^4 \, e^{- S_3/T},
\ee
where $A$ is a constant of ${\cal O}(1)$ and the Euclidean action is 
\be
S_3 = 4\pi \int d\varrho \, \varrho^2 \left[ \frac12 
\left(\frac{d\varphi}{d\varrho}\right)^2
+ V(\varphi,T)\right].
\ee
To determine the bounce configuration, we use the method presented in
Ref.~\cite{Konstandin:2006nd}. This method is based on a two step
procedure. First, the bounce solution is determined without damping,
$\gamma=0$, which can be achieved relatively simple due to energy
conservation. Subsequently, the parameter $\gamma$ is continuously
increased to the desired value, $\gamma=2$, while a linearized version
of the discretized equation of motion is iteratively solved.

The temperature when the phase transition occurs depends besides the
Euclidean action $S_3(T)$ on the cosmological
parameters~\cite{Anderson:1991zb}. The expansion of the Universe
is characterized by the Hubble parameter 
\be
H^{-1} = \frac{2 \xi M_{pl}}{T^2},
\ee
where $M_{pl}=1.22 \times 10^{19}$ GeV denotes the  Planck mass,
and near the electroweak phase transition $\xi \simeq 1/34$. 
The probability that a bubble was nucleated inside the causal volume
is given by
\be
\label{eq_dP}
dP =  A \frac{T^4}{H^4} e^{-S_3/T} \frac{dT}{T}.
\ee
The temperature of the beginning of the phase transition $T_n$ is
defined by the nucleation of the first bubble
\be
\label{eq_P}
\left. P \right|_{T=T_n} =  \int_{T_n}^\infty dP  = 1.
\ee
On the other hand, most gravitational radiation is emitted at the end
of the phase transition when the bubbles collide. Neglecting the
expansion of the Universe, the radius of a bubble at the temperature
$T_x$ that nucleated at temperature $T$ and expands with a velocity
$v_b$ is given by~\cite{Anderson:1991zb}
\be
\label{def_R}
R (T_x, T) = v_b \frac{T_x}{H(T_x)} 
\left( \frac1{T_x} - \frac1{T} \right).
\ee
Thus, the fraction of the causal volume that is in the broken phase is
\be
\label{eq_f}
f(T_x) \simeq \frac{4\pi H^3}{3 } \int_{T_x}^\infty R^3(T_x,T) dP. 
\ee
We define the end of the phase transition to be given by
\be
\label{def_Tf}
\left. f \right|_{T_x=T_f} = 1.
\ee
Another relevant quantity in the analysis of gravitational wave
production is the typical size of the nucleated bubbles at the end of
the phase transition, $\left< R \right>$. Following
Ref.~\cite{Nicolis:2003tg}, we use the maximum of the bubble volume
distribution, $dV \propto R^3 dP$, which should be closely related to
the typical scale for energy injection into the turbulent plasma. In
this case, one obtains the condition (see \eqs~(\ref{def_beta}) and
(\ref{def_R}))
\be
\label{def_R_details}
\left< R \right> \approx 3\frac{v_b}{\beta(T_R)}, \quad 
T_R = \frac{v_b T_f}{ v_b - H(T_f) \left< R \right>}.
\ee
This definition has the virtue that it characterizes typical bubble
sizes at the end of the phase transition in a reasonable way, even for
very strong phase transitions, in contrast to $\beta(T_f)$ that can
become negative. The temperature $T_R$ is by construction larger than
$T_f$ and typically smaller than $T_n$. Notice that there is a certain
arbitrariness in the choice of the length scale that enters into the
analysis of the peak frequency of the turbulent spectrum. The authors
of Ref.~\cite{Dolgov:2002ra} use the maximum of the bubble volume
distribution in momentum space, given by $\left< R
\right> \approx 5 v_b / \beta$, while the authors of
Ref.~\cite{Caprini:2006jb} prefer the size of the largest bubble
$\left< R \right> \approx v_b / \beta$. We employ the maximum of the
bubble volume distribution without transforming to momentum space what
leads to \eq~(\ref{def_R_details}) and compromises between these two
choices. This arbitrariness is a major source of uncertainty in the GW
spectrum contribution from turbulence. 

Finally, $N$ is the number of bubbles per causal volume at the end of
the phase transition
\be
\label{eq_N}
N = \left. P \right|_{T=T_f}.
\ee

To obtain approximate results for these quantities, a Taylor
expansion of the action can be used. Performing the integrals in
\eqs~(\ref{eq_P}), (\ref{eq_f}) and (\ref{eq_N}) leads to
\bea 
\label{Tn_approx}
\left. \frac{S_3}{T} \right|_{T=T_n} &\eqsim& 141.4  - 4 \log \left(
  \frac{T_n}{100 \textrm{GeV}}\right) - \log \left(
  \frac{\beta(T_n)}{100}\right), \\
\left. \frac{S_3}{T} \right|_{T=T_f} &\eqsim& 130.8  - 4 \log \left(
  \frac{T_f}{100 \textrm{GeV}}\right) - 
4 \log \left( \frac{\beta(T_f)}{100}\right) + 3 \log v_b, \\
\left. N \right|_{T=T_f} &\eqsim&  
\left. \frac{1}{8\pi } \left( \frac{\beta}{v_b \, H }\right)^3 \right|_{T=T_f}.
\eea     

\end{document}